\documentclass[aps,pra,10pt,twocolumn,superscriptaddress,nolongbibliography,nofootinbib]{revtex4-2}%
\pdfoutput=1%
\usepackage{graphicx}%
\graphicspath{{img/}}%
\usepackage{amsmath, amssymb}%
\usepackage{mathphyscmd}%
\everymath{\displaystyle}%
\usepackage{booktabs}%
\RequirePackage{siunitx}%
\usepackage{hyperref}%
\hypersetup{%
	pdfinfo={%
		Author       = {David Gaspard},
		Title        = {Effective-medium approach to the resonance distribution of wave scattering in a random point field},
		Subject      = {Waves in disordered media},
		CreationDate = {D:20230808132755+0200},
	},
	pdfborderstyle={/S/U/W 1},
	breaklinks=true
}
\makeatletter%
\g@addto@macro{\UrlBreaks}{\UrlOrds}%
\makeatother

\begin{document}%
\title{Effective-medium approach to the resonance distribution of wave scattering in a random point field}%

\author{David Gaspard}
\email[E-mail:~]{david.gaspard@espci.fr}
\affiliation{Nuclear Physics and Quantum Physics, CP229, Université libre de Bruxelles (ULB), B-1050 Brussels, Belgium}
\affiliation{Institut Langevin, ESPCI Paris, PSL University, CNRS, F-75005 Paris, France}

\author{Jean-Marc Sparenberg}
\affiliation{Nuclear Physics and Quantum Physics, CP229, Université libre de Bruxelles (ULB), B-1050 Brussels, Belgium}
\date{\today}

\begin{abstract}%
In a previous paper [\href{https://doi.org/10.1103/PhysRevA.105.042205}{Phys.\ Rev.\ A {\bf 105}, 042205 (2022)}], the distribution of resonance poles in the complex plane of the wavenumber $k$ associated to the multiple scattering of a quantum particle in a random point field was numerically discovered.
This distribution presented two distinctive structures: a set of peaks at small $k$ when the wavelength is larger than the interscatterer distance, and a band almost parallel to the real axis at larger $k$.
In this paper, a theoretical study based on wave transport theory is proposed to explain the origin of these structures and to predict their distribution in the complex $k$ plane.
First, it is shown that the peaks at small $k$ can be understood using the effective wave equation for the average wavefunction over the disorder.
Then, that the band at large $k$ can be described by the Bethe-Salpeter equation for the square modulus of the wavefunction.
This study is supported by careful comparisons with numerical simulations.
\end{abstract}%
\maketitle

\section{Introduction}\label{sec:intro}
The propagation of waves in disordered media is a subject of crucial importance in many research areas in physics such as acoustics \cite{Martin2018, Lanoy2015, Raveau2015, Wout2021, JerezBoudesseul2023}, electromagnetism \cite{Hsu2017, Binninger2019, Carminati2021, Monsarrat2022, Vynck2023}, and matter waves \cite{Kharchenko2001, Kuhn2007, Champenois2008, Billy2008, Miniatura2009, Jendrzejewski2012, Richard2019, Rose2022}.
These phenomena are described by wave equations with random or statistically correlated spatial heterogeneities in their propagation parameters such as the index of refraction or the potential energy function \cite{Barabanenkov1971, Barabanenkov1975, Bass1979, Rytov1989, Lagendijk1996, Rossum1999, Mishchenko2006, ShengP2006, Akkermans2007}.
Examples are given by multiple scattering of waves in a random field of scatterers such as atoms or molecules in gases, liquids, or amorphous solids.
Disorder is known to cause special effects on wave propagation, depending on the ratio of the wavelength over the scattering mean free path.
Pioneering works on multiple scattering by Foldy \cite{Foldy1945} and Lax \cite{Lax1951, *Lax1952} showed that a disordered medium may be considered under some conditions as an effective medium characterized by spatially averaged propagation parameters such as an effective refractive index.
Later on, Anderson discovered that waves may be localized as the result of multiple scattering in disordered media \cite{Anderson1958, Lagendijk2009}.
Otherwise, in semiclassical regimes, wave multiple scattering leads to transport ruled by the diffusion equation or the Boltzmann kinetic equation \cite{Chandrasekhar1960, Weinberg1958, Boltzmann2003}.
\par Disordered media may be infinite, semi-infinite, or finite with different geometries like spheres or cubes.
Much work is devoted to the study of wave propagation in infinite media or the transmission of waves through a slab, a wire, or a waveguide. 
Less is understood about the propagation and escape of waves from inside a finite disordered open system towards its exterior.
In weakly open systems such as a cavity connected to a few waveguides and terminals, the number of output channels is limited and the methods of random-matrix theory apply \cite{Beenakker1997, Kottos2005, Weiss2006, Skipetrov2011, Goetschy2011a, *Goetschy2011b, Fyodorov2015, Fyodorov2024}.
However, in many circumstances of great interest, the disordered system is finite and strongly open.
This is the case, in particular, for a fast quantum particle emitted by a source immersed in a gaseous detector such as a cloud chamber \cite{Sparenberg2018, GaspardD2019b}.
Here, a fundamental issue is to understand how the wavefunction of the emitted particle is affected by the disorder in the instantaneous configuration of the atoms composing the detector and how long the particle will travel inside the detector before escaping to the surrounding vacuum.
The characteristic times of this escape process may be determined in terms of the scattering resonances of the finite disordered medium formed by the random field of atoms composing the gaseous detector.
\par To simplify the huge complexity of the problem, this system can be described using the model of Foldy and Lax if the atoms of the gas can be represented by point-like scatterers in the sense that their radius is assumed to be much smaller than the particle wavelength in the $s$-wave approximation \cite{GaspardD2019b, GaspardD2022a, GaspardD2022b, Joachain2023}.
The advantage of this model is the reduction of the large multiple-scattering problem to a linear system, which can be solved numerically using modern computational resources to obtain the wavefunctions in great details.
Such models may be considered as quantum random Lorentz gases, as we did in our previous papers \cite{GaspardD2022a, GaspardD2022b}, where we carried out the numerical study of wave propagation in a spherical medium containing hundreds to thousands of fixed point scatterers for different values of their density with respect to the wavelength.
The differential and total cross sections of such systems were studied in Refs.\ \cite{GaspardD2022a, Joachain2023}.
Moreover, the distribution of the scattering resonances has also been investigated in the complex plane of the wavenumber using the resonance potential method we introduced in Ref.\ \cite{GaspardD2022b}.
In this way, we observed remarkable structures at low and high energies in the distribution of scattering resonances that were not understood in the numerical exploration carried out in our previous paper \cite{GaspardD2022b}.
\par The purpose of the present paper is to show that these structures in the resonance distribution can be quantitatively explained with accuracy using the effective-medium theory at complex values of the wavenumber.
The effective-medium wave equation, also referred to as the Dyson equation \cite{Barabanenkov1971, Barabanenkov1975, Bass1979, Rytov1989, ShengP2006, Akkermans2007, Doicu2018a1, *Doicu2019a2, *Doicu2019a3}, allows us to understand the formation of the low-energy structures previously observed in the resonance distribution.
Furthermore, we can show that the diffusion approximation to the transport equation, known in this research area as the Bethe-Salpeter equation \cite{Barabanenkov1971, Barabanenkov1975, Bass1979, Rytov1989, Mishchenko2006, ShengP2006, Akkermans2007, Fante1981, Fante1982, Tishkovets2006, Doicu2018a1, *Doicu2019a2, *Doicu2019a3}, can explain key features of the resonance distribution at high energies.
\par This paper is organized as follows.
The Foldy-Lax model used to describe the propagation of the quantum particle in a gaseous detector is presented in Sec.\ \ref{sec:foldy-lax}.
The scattering resonances in the complex plane of the wavenumber are defined in Sec.\ \ref{sec:scattering-resonances}.
The disorder-averaged Green function is introduced in Sec.\ \ref{sec:avg-green}.
This function is at the heart of the semiclassical multiple-scattering theory for the square modulus of the Green function developed in Sec.\ \ref{sec:semiclassical-transport}.
Both disorder-averaging approaches are used to explain the structures observed in the resonance distribution.
The predictions of the theory are then compared in Sec.\ \ref{sec:numerical-results} with the exact resonance distributions computed numerically.
Finally, conclusions are drawn in Sec.\ \ref{sec:conclusions}.
\par Since the models considered in this paper are valid in a space of arbitrary dimension, the volume and surface area of the unit ball in the space $\mathbb{R}^d$ frequently appear.
They are respectively given by
\begin{equation}\label{eq:ball-surf-vol}
V_d = \frac{\pi^{\frac{d}{2}}}{\Gamma(\frac{d}{2}+1)} \qquad\text{and}\qquad
S_d = dV_d = \frac{2\pi^{\frac{d}{2}}}{\Gamma(\frac{d}{2})}  \:,
\end{equation}
where $\Gamma(z)$ denotes the gamma function \cite{Olver2010}.
\par All the numerical results presented in this paper are computed with the program MSModel \cite{GaspardD2023-prog}.

\section{Wave scattering in random media}\label{sec:waves-in-random-media}

\subsection{The Foldy-Lax model}\label{sec:foldy-lax}
In order to describe the multiple collisions of the quantum particle in a disordered medium, we consider a simple but powerful model originally introduced by Foldy and Lax \cite{Foldy1945, Lax1951, *Lax1952, Mishchenko2006}.
In this model, the wavefunction of the particle undergoes elastic collisions without loss of energy in a random configuration of fixed point scatterers representing the atoms or the molecules of a gaseous detector.
This model is very convenient especially because it can be efficiently solved numerically for large disordered systems without resorting to huge discretization lattices. %
In this model, the wavefunction $\psi(\vect{r})$ obeys the stationary Schrödinger equation
\begin{equation}\label{eq:schrodinger-general}
\left[ \lapl_{\vect{r}} + k^2 - U(\vect{r}) \right] \psi(\vect{r}) = 0  \:,
\end{equation}
where $k=2\pi/\lambda$ is the wavenumber and $\lambda$ the wavelength.
The potential $U(\vect{r})$ in Eq.\ \eqref{eq:schrodinger-general} reads
\begin{equation}\label{eq:general-multi-potential}
U(\vect{r}) = \sum_{i=1}^N u(\vect{r}-\vect{x}_i)  \:,
\end{equation}
where $u(\vect{r})$ is the short-range potential of the scatterers that we will define in details soon.
\par In Eq.\ \eqref{eq:general-multi-potential}, the scatterers are located at the fixed positions $\vect{x}_1,\vect{x}_2,\ldots,\vect{x}_N$.
This random point field is characterized by the joint probability distribution $P(\vect{x}_1,\ldots,\vect{x}_N)$.
In this regard, the statistical average of some observable $A$ over the random configurations of the scatterers is given by
\begin{equation}\label{eq:config-average}\begin{split}
\avg{A(\vect{r})} & = \int_{\mathbb{R}^{d}}\D\vect{x}_1 \cdots \int_{\mathbb{R}^{d}}\D\vect{x}_N  \\
 & A(\vect{r};\vect{x}_1,\ldots,\vect{x}_N) \,P(\vect{x}_1,\ldots,\vect{x}_N)  \:.
\end{split}\end{equation}
In particular, we define the local average density of scatterers per unit volume as
\begin{equation}\label{eq:local-medium-density}
n(\vect{r}) = \avg{\sum_{i=1}^N \delta(\vect{r}-\vect{x}_i)}  \:.
\end{equation}
Unless otherwise stated, $\tavg{\cdots}$ will refer to the configurational average \eqref{eq:config-average}.
Furthermore, we assume that the positions $\vect{x}_1,\ldots,\vect{x}_N$ are independent and identically distributed random variables, so that $P(\vect{x}_1,\ldots,\vect{x}_N)$ factorizes into $P(\vect{x}_1)\cdots P(\vect{x}_N)$.
In addition, we assume that the scatterers are contained in region $\mathcal{V}$, and that the distribution is uniform. Therefore, the scatterer density reads
\begin{equation}\label{eq:uniform-density}
n(\vect{r}) = \begin{cases}%
N/V  & \text{if}~\vect{r}\in\mathcal{V}  \:,\\
0    & \text{otherwise}  \:,
\end{cases}\end{equation}
where $V$ is the volume of $\mathcal{V}$.
In all numerical simulations, we will use the mean interparticle distance
\begin{equation}\label{eq:def-unit-length}
\varsigma = \left(\frac{V}{N}\right)^{\frac{1}{d}}  \:,
\end{equation}
as the unit length.
\par In Eq.\ \eqref{eq:general-multi-potential}, the potential $u(\vect{r})$ is supposed to have a finite spatial range $b$ much smaller than the wavelength ($kb\ll 1$), so that the point scattering theory developed in Sec.\ III of Ref.\ \cite{GaspardD2022a} can be applied.
The short-range potential $u(\vect{r})$ is actually defined such that the solution of Eq.\ \eqref{eq:schrodinger-general} for $N=1$ reads
\begin{equation}\label{eq:wavefun-one-point}
\psi(\vect{r}) = \phi(\vect{r}) + F(k) \gfree^+(k,\vect{r}\mid\vect{x}_1)  \:,
\end{equation}
where $\phi(\vect{r})=\E^{\I\vect{k}\cdot\vect{r}}$ is the incident plane wave, and $F(k)$ is the point scattering amplitude.
The quantity $F(k)$ is an intrinsic property of the scatterer and therefore is not affected by the presence of other scatterers.
We will explain more about $F(k)$ soon.
\par In Eq.\ \eqref{eq:wavefun-one-point}, $\gfree(k,\vect{r}\mid\vect{r}')$ is the free Green function which satisfies
\begin{equation}\label{eq:def-free-green}
\left(\lapl_{\vect{r}} + k^2\right) \gfree(k,\vect{r}\mid\vect{r}') = \delta(\vect{r}-\vect{r}')  \:.
\end{equation}
It is worth noting that the wavenumber $k$ in Eq.\ \eqref{eq:def-free-green} is a complex number ($k=\freal{k}+\I\fimag{k}$).
This complex nature plays an important role in this paper.
The solution of Eq.\ \eqref{eq:def-free-green} reads
\begin{equation}\label{eq:free-green-bessel-k}
\gfree^\pm(k,r) = -\frac{1}{2\pi} \left(\frac{\mp\I k}{2\pi r}\right)^{\frac{d-2}{2}} K_{\frac{d-2}{2}}(\mp\I kr)  \:,
\end{equation}
where $K_\nu(z)$ is the modified Bessel function \cite{Olver2010} and $r=\norm{\vect{r}-\vect{r}'}$.
The free Green function \eqref{eq:free-green-bessel-k} is characterized by the asymptotic behavior
\begin{equation}\label{eq:free-green-asym}
\gfree^\pm(k,r) \xrightarrow{r\rightarrow\infty} \pm\frac{1}{2\I k}\left(\frac{\mp\I k}{2\pi r}\right)^{\frac{d-1}{2}} \E^{\pm\I kr} \:.
\end{equation}
\par The symbol $F(k)$ in Eq.\ \eqref{eq:wavefun-one-point} stands for the point scattering amplitude \cite{GaspardD2022a}
\begin{equation}\label{eq:point-ampli-from-phase-shift}
F(k) = \frac{1}{\pi\dos(k) (\I - \cot\delta(k))}  \:.
\end{equation}
In Eq.\ \eqref{eq:point-ampli-from-phase-shift}, $\delta(k)$ is the $s$-wave phase shift \cite{Joachain1979, Newton1982, Taylor2006, Sakurai2020} and $\dos(k)$ is the local density of states per unit of $k^2$ in free space given by
\begin{equation}\label{eq:def-free-dos}
\dos(k) = \bra{\vect{r}} \delta(k^2 - \op{\vect{p}}^2) \ket{\vect{r}} = \frac{S_d k^{d-2}}{2(2\pi)^d}  \:,
\end{equation}
where $\op{\vect{p}}=-\I\grad_{\vect{r}}$ is the momentum operator.
The cross section of the point scatterer is related to the scattering amplitude $F(k)$ by \cite{GaspardD2022a}
\begin{equation}\label{eq:point-total-cross-section}
\sigma_{\rm pt}(k) = \frac{\pi}{k} \dos(k) \abs{F(k)}^2  \:.
\end{equation}
The phase shift in the point scattering amplitude \eqref{eq:point-ampli-from-phase-shift} intrinsically depends on the microscopic model used to describe the scattering over the individual scatterers.
When the radius of the individual scatterers is much smaller than the wavelength, the scattering amplitude universally behaves as \cite{GaspardD2022a}
\begin{equation}\label{eq:point-ampli-hard-sphere}
F_{\rm hs}(k) = -\frac{I(k,\alpha)}{I(k,0)\gfree^+(k,\alpha)}  \:,
\end{equation}
where $I(k,r)=-\Im\gfree^+(k,r)$ and $\alpha$ is known as the scattering length.
This quantity must be smaller than the wavelength for this model to be valid ($\alpha k\ll 1$).
Beside this, the point cross section \eqref{eq:point-total-cross-section} can be maximized by imposing $\cot\delta(k)=0$ in Eq.\ \eqref{eq:point-ampli-from-phase-shift}.
This provides the parameter-free scattering model:
\begin{equation}\label{eq:point-ampli-max}
F_{\max}(k) = \frac{1}{\I\pi\dos(k)}  \:.
\end{equation}
The point scattering model \eqref{eq:point-ampli-max} has the particularity of saturating the upper bound for the point cross section \cite{GaspardD2022a}
\begin{equation}\label{eq:point-cross-section-upper-bound}
\sigma_{\max}(k) = \frac{1}{\pi k\dos(k)}  \:.
\end{equation}
Equation \eqref{eq:point-ampli-max} thus produces the largest cross section possible for any given value of $k$.
On the one hand, this is useful to effectively exploit each individual scatterer, especially in small point fields ($N\lesssim 10^2$).
On the other hand, this model should be used with caution at small values of $k$ because the cross section \eqref{eq:point-cross-section-upper-bound} diverges as $\bigo(k^{1-d})$ in dimensions $d\geq 2$ in a non-physical way.
Another feature of the model \eqref{eq:point-ampli-max} is the absence of resonance corresponding to the single-point scattering.
\par We have now all the ingredients to solve the full wave equation \eqref{eq:schrodinger-general}.
Given the form of the potential \eqref{eq:general-multi-potential}, the particle wavefunction $\psi(\vect{r})$ should read
\begin{equation}\label{eq:multi-wave-function}
\psi(\vect{r}) = \phi(\vect{r}) + \sum_{i=1}^N a_i G^+(k,\vect{r}\mid\vect{x}_i)  \:,
\end{equation}
The wave amplitudes $a_i~\forall i\in\{1,\ldots,N\}$ in Eq.\ \eqref{eq:multi-wave-function} satisfy the self-consistent equation
\begin{equation}\label{eq:multi-lippmann-schwinger}
a_i = F(k)\left(\phi(\vect{x}_i) + \sum_{j (\neq i)}^N a_j G^+(k,\vect{x}_i\mid\vect{x}_j)\right)  \:.
\end{equation}
Equation \eqref{eq:multi-lippmann-schwinger} can be rewritten in matrix form using the vector notations $\vect{\phi}=\tran{\left(\phi(\vect{x}_1),\phi(\vect{x}_2),\ldots,\phi(\vect{x}_N)\right)}$ and $\vect{a}=\tran{\left(a_1,a_2,\ldots,a_N\right)}$. The result is
\begin{equation}\label{eq:m-matrix-lippmann-schwinger}
\matr{M}(k)\,\vect{a} = \vect{\phi}  \:,
\end{equation}
where the $N\times N$ multiple-scattering matrix $\matr{M}(k)$ is defined by
\begin{equation}\label{eq:def-m-matrix}
\matr{M}(k) = F(k)^{-1} \matr{1} - \matr{G}^+(k)  \:,
\end{equation}
and $\matr{1}$ is the identity matrix.
The free Green matrix $\matr{G}^+(k)$ in Eq.\ \eqref{eq:def-m-matrix} is given by
\begin{equation}\label{eq:def-green-matrix}
G^+_{ij}(k) = G^+(k,\vect{x}_i\mid\vect{x}_j)(1-\delta_{ij})  \:.
\end{equation}
The wavefunction is thus obtained from Eq.\ \eqref{eq:multi-wave-function} where the coefficients $a_i=[\matr{M}(k)^{-1}\vect{\phi}]_i$ are given by the inversion of the matrix \eqref{eq:def-m-matrix}.
The solution of the linear system \eqref{eq:m-matrix-lippmann-schwinger} and the corresponding wavefunction \eqref{eq:multi-wave-function} can be computed using the program MSModel \cite{GaspardD2023-prog}.
\par
\begin{figure*}[ht]%
\includegraphics{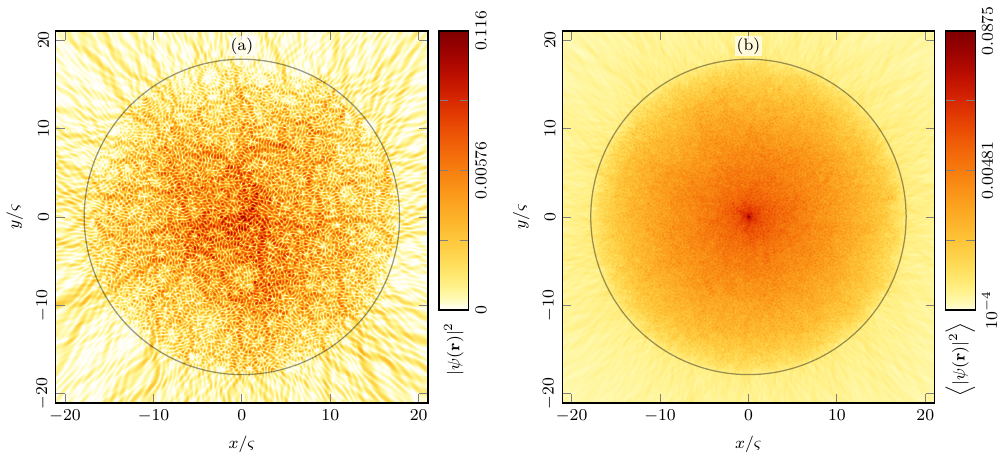}%
\caption{(a) Square modulus of the wavefunction, $\abs{\psi(\vect{r})}^2$, computed from Eqs.\ \eqref{eq:multi-wave-function}, \eqref{eq:m-matrix-lippmann-schwinger}, and \eqref{eq:incident-spherical-wave} using the program MSModel \cite{GaspardD2023-prog} in a 2D disk-shaped random point field of $N=10^{3}$ scatterers with $k=10\,\varsigma^{-1}$.
The point scattering model is Eq.\ \eqref{eq:point-ampli-max} and the mean free path is thus $\lscat=2.5\,\varsigma$.
The circle depicts the medium boundary.
(b) Same settings as panel~(a) but averaged over $64$ disorder configurations.}%
\label{fig:wfun-density-spherical}%
\end{figure*}

 An example of the square modulus of the wavefunction for a large random configuration of $N=10^3$ scatterers located in a two-dimensional (2D) disk-shaped region is shown in Fig.\ \ref{fig:wfun-density-spherical}(a) using the incident wave
\begin{equation}\label{eq:incident-spherical-wave}
\phi(\vect{r}) = G^+(k,\vect{r}\mid\vect{r}_0)  \:.
\end{equation}
The incident wave \eqref{eq:incident-spherical-wave} represents a quantum particle created at the center of the medium ($\vect{r}_0=\vect{0}$).
Figures similar to Fig.\ \ref{fig:wfun-density-spherical} exist in the literature for other systems \cite{Vanneste2005}.
The average of the square modulus of the wave function over several random configurations of the scatterers is shown in panel \ref{fig:wfun-density-spherical}(b).
This panel illustrates in particular that disorder averaging restores the spherical symmetry of the problem due to the uniformity of the scatterer density \eqref{eq:uniform-density}.

\subsection{Scattering resonances}\label{sec:scattering-resonances}
This paper aims at studying the distribution of resonances which are formally defined as the singular points, or poles, of the resolvent $\op{\gfull}(k)$ of the complete multiple-scattering problem \eqref{eq:schrodinger-general}--\eqref{eq:general-multi-potential}.
This resolvent operator is defined by
\begin{equation}\label{eq:def-full-resolvent}
\op{\gfull}(k) = \frac{1}{k^2 - \op{\vect{p}}^2 - U(\op{\vect{r}})}  \quad\forall k\in\mathbb{C}\setminus\mathbb{R}  \:,
\end{equation}
and is related to the full Green function by $\gfull(k,\vect{r}\mid\vect{r}')=\bra{\vect{r}}\op{\gfull}(k)\ket{\vect{r}'}$.
The latter function describes the propagation from $\vect{r}'$ to $\vect{r}$ in presence of the random point field.
The poles of Eq.\ \eqref{eq:def-full-resolvent} on the positive imaginary semiaxis of the wavenumber ($\Im k>0$) correspond to the bound states of the system.
The resolvent \eqref{eq:def-full-resolvent} also displays a continuum of poles, also known as a branch cut, on the real axis of $k$ which is interpreted as the continuum of (unbound) scattering states.
To investigate on the characteristics of the scattering states, in particular to study the time required for the quantum particle to escape the system, it is possible to achieve an \emph{analytic continuation} through this branch cut.
In general, this continuation reveals other poles in the vicinity of the real axis of $k$, but in the lower half-plane ($\Im k<0$), which are interpreted as resonant states.
The imaginary part of these poles gives the escape rate of the particle from the system.
Indeed, if we assume that the wavefunction behaves as $\psi(t)\propto\E^{-\I\omega t}$, then the square modulus will decay as $\abs{\psi(t)}^2=\E^{2\Im(\omega)t}=\E^{-\Gamma t}$.
Therefore, we identify the escape rate as
\begin{equation}\label{eq:def-escape-rate}
\Gamma = -2\Im[\omega(\freal{k} + \I\fimag{k})] = -2v(\freal{k}) \fimag{k} + \bigo(\fimag{k}^3)  \:,
\end{equation}
where $k=\freal{k}+\I\fimag{k}$ is the position of the resonance pole, and $v(k)=\partial_k\omega(k)$ is the group velocity.
Since $\fimag{k}<0$ for a resonance pole, the escape rate \eqref{eq:def-escape-rate} is positive.
\par The advantage of the Foldy-Lax model is to provide a systematic way of defining the complex resonance poles associated to the multiple scattering of the wave in the random point field.
Indeed, using the multiple-scattering matrix \eqref{eq:def-m-matrix}, the full resolvent \eqref{eq:def-full-resolvent} reads
\begin{equation}\label{eq:full-resolvent-multi-1}
\op{\gfull}(k) = \op{G}(k) + \sum_{i,j}^N \op{G}(k)\ket{\vect{x}_i} [\matr{M}(k)^{-1}]_{ij} \bra{\vect{x}_j}\op{G}(k)  \:.
\end{equation}
Due to Eq.\ \eqref{eq:full-resolvent-multi-1}, the resonance poles which interest us are the solutions of the determinantal equation
\begin{equation}\label{eq:determinantal-equation}
\det\matr{M}(k) = 0  \quad\text{for}~k\in\mathbb{C}  \:.
\end{equation}
Equation \eqref{eq:determinantal-equation} has infinitely many roots in the complex plane of $k$, and the amount of these roots increases with the number $N$ of scatterers.
We thus define the complex resonance density as \cite{GaspardD2022b}
\begin{equation}\label{eq:def-resonance-density}
\docres(\freal{k},\fimag{k}) = \frac{1}{N} \sum_{p=1}^{\infty} \avg{\delta^{(2)}(k - k_p)}  \:,
\end{equation}
where the sum is over the resonance poles and $\delta^{(2)}(k-k_p)$ is the two-dimensional Dirac delta over the complex plane of $k$.
The two-dimensional distribution \eqref{eq:def-resonance-density} can be obtained numerically with the resonance potential method \cite{GaspardD2022b} which consists in the evaluation of
\begin{equation}\label{eq:resonance-density-from-pot}
\docres(\freal{k},\fimag{k}) = \frac{1}{2\pi N} \left( \pder[2]{}{\freal{k}} + \pder[2]{}{\fimag{k}} \right) \avg{\ln\abs{\det\matr{M}(k)}}  \:.
\end{equation}
This average resonance distribution displays the structures seen, for instance, in Fig.\ 6 of Ref.\ \cite{GaspardD2022b}, and is the focus of interest in this paper.

\subsection{Disorder-averaged Green function}\label{sec:avg-green}
In this section, we develop a theory for the low-energy peaks in the complex resonance density observed in our previous paper \cite{GaspardD2022b}.
These peaks appear at low energy, that is in a regime of large wavelength compared to the typical interscatterer distance ($k\varsigma\ll 1$).
In this regime, we expect that the wave is no longer able to resolve the spatial heterogeneities of the medium, and thus perceives a uniform \emph{effective medium} \cite{ShengP2006}.
A possible approach to derive an effective-medium wave equation is to average directly the wavefunction over the random configurations of the scatterers, hence leading to a Dyson-type wave equation \cite{Barabanenkov1971, Barabanenkov1975, Bass1979, ShengP2006, Akkermans2007, Doicu2018a1, *Doicu2019a2, *Doicu2019a3}.
As we will see, this equation successfully predicts the locations of the resonance peaks at low energy, providing a quantitative understanding of the numerical observations in our previous paper \cite{GaspardD2022b}.
\par In wave transport theory, an important quantity is the full Green function averaged over the realizations of the disorder.
It can be shown that the average Green function $\tavg{\gfull(k,\vect{r}\mid\vect{r}')}$ obeys the Dyson-type effective-medium equation \cite{Foldy1945, Lax1951, *Lax1952, Barabanenkov1971, Barabanenkov1975, Bass1979, ShengP2006, Akkermans2007, Doicu2018a1, *Doicu2019a2, *Doicu2019a3, GaspardD2022-thesis, GaspardD2023-arxiv-v1}
\begin{equation}\label{eq:avg-green-schrodinger}
\left[ \lapl_{\vect{r}} + k^2 - n(\vect{r})F(k) \right] \avg{\gfull(k,\vect{r}\mid\vect{r}')} = \delta(\vect{r}-\vect{r}')  \:,
\end{equation}
where $n(\vect{r})$ is the scatterer density \eqref{eq:local-medium-density} and $F(k)$ is the point scattering amplitude \eqref{eq:point-ampli-from-phase-shift}.
Equation \eqref{eq:avg-green-schrodinger} is very close to the Green-function equation associated to Eq.\ \eqref{eq:schrodinger-general} but with the potential $U(\vect{r})$ replaced by $n(\vect{r})F(k)$.
By analogy with Eq.\ \eqref{eq:def-free-green}, the average Green function has two solutions valid for $k\in\mathbb{C}$: $\tavg{\gfull^+(k,\vect{r}\mid\vect{r}')}$ which exponentially vanishes for $\norm{\vect{r}-\vect{r}'}\rightarrow\infty$ on the domain $\Im k>0$, and $\tavg{\gfull^-(k,\vect{r}\mid\vect{r}')}$ which exponentially increases on the same domain.
\par Often, it is convenient to introduce the (local) effective wavenumber \cite{Foldy1945, Lax1951, Twersky1964, ShengP2006, Akkermans2007}
\begin{equation}\label{eq:effective-wavenumber}
\kappa(k,\vect{r}) = \sqrt{k^2 - n(\vect{r})F(k)}  \:,
\end{equation}
such that, in regions of constant $n(\vect{r})$, the average Green function essentially behaves as the free Green function
\begin{equation}\label{eq:avg-green-sol}
\avg{\gfull^+(k, r)} = G^+(\kappa(k), r)  \:,
\end{equation}
where $r=\norm{\vect{r}-\vect{r}'}$. 
Due to the complex nature of the point scattering amplitude $F(k)$, the effective wavenumber $\kappa$ is itself complex, meaning that the average Green function exponentially decays with the distance $r$ from the point source, even for $k\in\mathbb{R}$ because of disorder.
This decay can be highlighted using the asymptotic behavior \eqref{eq:free-green-asym}:
\begin{equation}\label{eq:avg-green-behavior-1}
\abs{\avg{\gfull^+(k,r)}}^2 \xrightarrow{r\rightarrow\infty} \frac{\pi\dos(\abs{\kappa})}{\abs{\kappa}S_dr^{d-1}} \E^{-2\Im\kappa r}  \:.
\end{equation}
The imaginary part of $\kappa$, given by Eq.\ \eqref{eq:effective-wavenumber} with $n(\vect{r})=n$, can be expanded if we assume that $k^2$ is relatively large compared to $nF(k)$.
We find
\begin{equation}\label{eq:imag-kappa-expansion}\begin{split}
2\Im\kappa(k) & = 2\Im\left(k - \frac{n}{2k}F(k) + \bigo\!\left[\tfrac{1}{k^3}(nF)^2\right]\right)  \\
 & \simeq 2\fimag{k} - \frac{n}{k}\Im F(k)  \:.
\end{split}\end{equation}
In virtue of the optical theorem, given for instance by Eq.\ (44) of Ref.\ \cite{GaspardD2022a}, it turns out that the last term of Eq.\ \eqref{eq:imag-kappa-expansion} is just the inverse of the scattering \emph{mean free path} \cite{ShengP2006, Akkermans2007, Cherroret2009-thesis}
\begin{equation}\label{eq:lscat-from-cross-section}
\frac{1}{\lscat} = n\sigma = -\frac{n}{k}\Im F(k) \:.
\end{equation}
The characteristic length $\lscat$ represents the mean distance between two successive collisions of the wave in the medium and thus plays a crucial role in transport theory.
This decay represents the loss of coherence of the incident wave due to the disorder.
Another contribution to the decay in Eq.\ \eqref{eq:imag-kappa-expansion} comes from the intrinsic imaginary part of $k$ which is essential for the study of complex resonances as we saw in Eq.\ \eqref{eq:def-escape-rate}.
In anticipation of future calculations, we already introduce the notation
\begin{equation}\label{eq:def-gamma-from-k}
\gamma = 2\fimag{k}  \:,
\end{equation}
such that the behavior \eqref{eq:avg-green-behavior-1} of the average Green function reads
\begin{equation}\label{eq:avg-green-behavior-2}
\abs{\avg{\gfull^+(k,r)}}^2 \xrightarrow{r\rightarrow\infty} \frac{\pi\dos(\abs{\kappa})}{\abs{\kappa}S_dr^{d-1}} \E^{-(\gamma+n\sigma)r}  \:.
\end{equation}
We will see in Sec.\ \ref{sec:low-energy-resonances} that the effective-medium equation \eqref{eq:avg-green-schrodinger} can be used to predict the peaks in the resonance density in the low-energy region. %

\subsection{Semiclassical transport}\label{sec:semiclassical-transport}
The issue with the effective-medium equation \eqref{eq:avg-green-schrodinger} is that, due to the destructive interference caused by the disorder
averaging, the wave undergoes exponential damping at the scale of the mean free path, alike in absorption processes. %
Therefore, this equation is not adapted to describe the multiple scattering which takes place on scales larger than one mean free path and which is responsible in particular for the spatial diffusion of the wave intensity.
We may expect that this multiple scattering will significantly affect the distribution of the complex resonances since their location is governed by the characteristic escape rate from the disordered region.
To address the issue of multiple scattering, we need to consider a more elaborate theory for wave transport that is referred to as the \emph{semiclassical theory} \cite{Berry1972, Nussenzveig1992}.
The central quantity of this theory is the intensity of the full Green function
\begin{equation}\label{eq:def-density}
\rho(\vect{r},\gamma) = \abs{ \gfull^+(k_0 + \tfrac{\I}{2}\gamma, \vect{r}\mid\vect{r}_0) }^2  \:,
\end{equation}
where $\vect{r}_0$ is the source point of the wave.
Note that $\rho(\vect{r},\gamma)$ fluctuates with the disorder as seen in Fig.\ \ref{fig:wfun-density-spherical}(a) but possesses a well-defined average shown in Fig.\ \ref{fig:wfun-density-spherical}(b).
The equation governing the disorder-averaged intensity, $\tavg{\rho(\vect{r},\gamma)}$, is known as the \emph{Bethe-Salpeter equation} and reads \cite{Foldy1945, Bourret1962, Barabanenkov1971, Barabanenkov1975, Bass1979, Rossum1999, ShengP2006, Akkermans2007, Kupriyanov2017, GaspardD2022-thesis, GaspardD2023-arxiv-v1}
\begin{equation}\label{eq:bethe-salpeter-diag}
\avg{\rho(\vect{r})} = K(\vect{r}\mid\vect{r}_0) + \int_{\mathbb{R}^d} \D{\vect{r}'} K(\vect{r}\mid\vect{r}') n(\vect{r}') \sigma(k) \avg{\rho(\vect{r}')}  \:,
\end{equation}
where the transport kernel $K(\vect{r}\mid\vect{r}')$ is defined by the square modulus of the effective Green function
\begin{equation}\label{eq:def-transport-kernel}
K(\vect{r}\mid\vect{r}') = \frac{\abs{k}}{\pi\dos(\abs{k})} \abs{\avg{\gfull^+(k,\vect{r}\mid\vect{r}')}}^2  \:.
\end{equation}
The integral kernel $K(\vect{r}\mid\vect{r}')$ can be interpreted as the probability density for the next collision point $\vect{r}$ given the previous collision happened at $\vect{r}'$.
The usefulness of the prefactor in Eq.\ \eqref{eq:def-transport-kernel} is that the far-field behavior of $K(\vect{r}\mid\vect{r}')$ does not explicitly depend on the wavenumber $k$.
In the weak scattering regime ($\abs{\kappa}\simeq\abs{k}$), one finds from Eq.\ \eqref{eq:avg-green-behavior-2}
\begin{equation}\label{eq:transport-kernel-asym}
K(\vect{r}\mid\vect{r}') \xrightarrow{r\rightarrow\infty} \frac{\E^{-(\gamma+n\sigma)\norm{\vect{r}-\vect{r}'}}}{S_d\norm{\vect{r}-\vect{r}'}^{d-1}}  \:.
\end{equation}
Note that the small imaginary part of $k=k_0+\I\fimag{k}$ can be neglected in the cross section because the equation only considers time scales much longer than the oscillation period of the wave ($\fimag{k}\ll k_0$). It is thus safe to write $\sigma(k)\simeq\sigma(k_0)$.
\par The main assumption behind the Bethe-Salpeter equation \eqref{eq:bethe-salpeter-diag} is that the wavelength is much smaller than the mean free path\footnote{Note that, since the mean free path is at least equal to the mean interatomic distance ($\lscat\geq\varsigma$), the condition \eqref{eq:weak-scattering} is necessarily satisfied for $k_0\varsigma\gg 1$.}
\begin{equation}\label{eq:weak-scattering}
k_0\lscat \gg 1  \:.
\end{equation}
The condition \eqref{eq:weak-scattering} is known in the literature as the \emph{weak scattering regime} \cite{ShengP2006}, or the \emph{weak disorder regime} \cite{Akkermans2007}.
In this regime, coherent effects such as localization play less significant roles, especially in 2D and 3D. %
Note that, in the context of gaseous particle detectors, the condition \eqref{eq:weak-scattering} is usually met.
In this regard, some orders of magnitude are given in Tab.\ \ref{tab:mean-free-path}.
\begin{table}[ht]%
\centering\renewcommand{\arraystretch}{1.3}%
\begin{tabular}{*{5}{l}}%
\toprule
Particle  & $k_0$ & $\lscat$ & $k_0\lscat$  \\
\midrule
Electron, $\SI{10}{\eV}$ \cite{Plante2009, Kennerly1980}  & \SI{16}{\per\nano\meter}   & \SI{0.4}{\micro\meter} & \num{6e3}  \\
Alpha, $\SI{5}{\MeV}$ \cite{Rutherford1924}               & \SI{1}{\per\femto\meter}   & \SI{1}{\micro\meter}   & \num{1e9}  \\
Photon, $\SI{550}{\nano\meter}$ \cite{Sneep2005}          & \SI{11}{\per\micro\meter}  & \SI{100}{\kilo\meter}  & \num{1e12} \\
\bottomrule
\end{tabular}%
\caption{Orders of magnitude for the wavenumber and the mean free path of some particles in dry ambient air ($P=\SI{e5}{\pascal}$ and $T=\SI{273}{\kelvin}$).
The molecular density is $n=\SI{0.0265}{\per\cubic\nano\meter}$. The visible light data only assumes the Rayleigh scattering.}%
\label{tab:mean-free-path}%
\end{table}%
Therefore, we find reasonable to neglect these effects in this paper.
\par The formal solution of the transport equation \eqref{eq:bethe-salpeter-diag} is given by
\begin{equation}\label{eq:bethe-salpeter-formal-sol}
\avg{\rho(\op{\vect{r}})} = \frac{1}{1 - \op{K} n(\op{\vect{r}})\sigma(k)} K(\op{\vect{r}}\mid\vect{r}_0)   \:.
\end{equation}
As we will see in Sec.\ \ref{sec:numerical-results}, the integral equation \eqref{eq:bethe-salpeter-diag} provides a prediction of the position of the peaks in the resonance density observed in our previous paper \cite{GaspardD2022b}.
According to Eqs.\ \eqref{eq:def-density} and \eqref{eq:bethe-salpeter-formal-sol}, we expect that most of the complex resonances occur in the vicinity of the singularities of $\tavg{\rho(\vect{r},\gamma)}$ in the variable $\gamma$.
These singularities are given by the eigenvalue problem
\begin{equation}\label{eq:transport-eigenvalue-system}
\avg{\rho(\vect{r})} = \int_{\mathbb{R}^d} \D{\vect{r}'} K(\vect{r}\mid\vect{r}') n(\vect{r}')\sigma(k) \avg{\rho(\vect{r}')}  \:,
\end{equation}
where the free variable $\gamma$ (defined in Eq.\ \eqref{eq:def-gamma-from-k} and used in Eq.\ \eqref{eq:transport-kernel-asym}) plays the role of the sought eigenvalue and the density $\avg{\rho(\vect{r})}$ is the associated eigenfunction.
The singular values $\gamma$ given by Eq.\ \eqref{eq:transport-eigenvalue-system} are expected to correspond to the main structures observed in the resonance density $\docres(\freal{k},\fimag{k})$ in the high-energy regime where the coherent (phase-dependent) effects are negligible.

\section{Numerical results}\label{sec:numerical-results}
In this section, the predictions of the effective-medium equations of Secs.\ \ref{sec:avg-green} and \ref{sec:semiclassical-transport} are compared to the exact density of complex resonances obtained numerically using the resonance potential method developed in our previous paper \cite{GaspardD2022b}.
In that paper, two distinctive structures are observed in the complex resonance spectrum: 
\begin{enumerate}
\item Peaks at low energy, that is in the region
\begin{equation}\label{eq:def-low-energy}
\abs{k}\varsigma \lesssim j_{\frac{d-2}{2}}  \:,
\end{equation}
where $\varsigma$ is the mean interscatterer distance \eqref{eq:def-unit-length} and $j_{\frac{d-2}{2}}$ is the first zero of the Bessel function $J_{\frac{d-2}{2}}(z)$.
These peaks are explained in Sec.\ \ref{sec:low-energy-resonances} by the Dyson-type effective-medium equation \eqref{eq:avg-green-schrodinger}.
\item A nearly horizontal resonance band at larger energies which is characterized in Sec.\ \ref{sec:resonance-band} using the Bethe-Salpeter equation of the form \eqref{eq:transport-eigenvalue-system}.
\end{enumerate}

\subsection{Low-energy peaks in the resonance density}\label{sec:low-energy-resonances}
\begin{figure*}[ht]%
\includegraphics{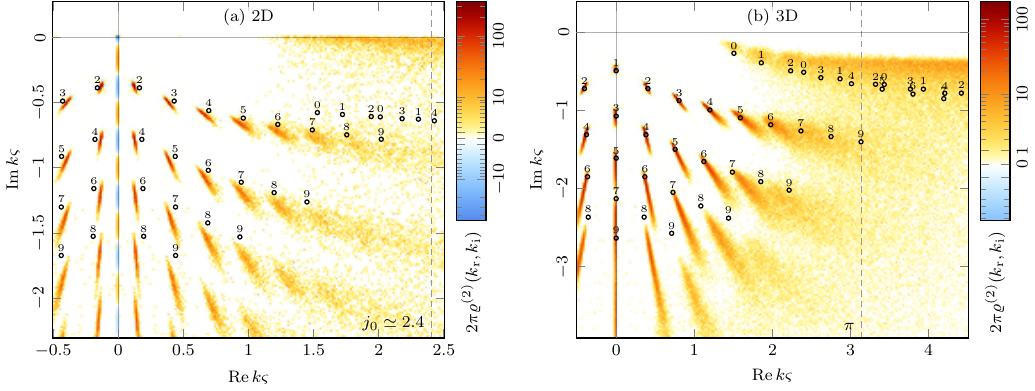}%
\caption{(a) Resonance density \eqref{eq:def-resonance-density} computed with the program MSModel \cite{GaspardD2023-prog} in a 2D disk-shaped random point field for $N=50$ and using the point scattering model \eqref{eq:point-ampli-hard-sphere} with $\alpha=0.1\,\varsigma$.
The parameters are thus $R=3.99\,\varsigma$ and $\lscat\simeq 1\,\varsigma$ (multiple-scattering regime).
The density is averaged over $2^8$ disorder configurations.
The resonances of the effective medium, obtained by numerically solving Eq.\ \eqref{eq:eff-medium-resonance-2}, are depicted by small circles labelled with the corresponding angular momentum $\ell$.
The dashed line demarcates the low-energy region \eqref{eq:def-low-energy}.
(b) Same as panel~(a) but in a 3D spherical random point field for $N=100$. %
The parameters are thus $R=2.88\,\varsigma$ and $\lscat\simeq 10\,\varsigma$ (quasiballistic regime).}%
\label{fig:k-plane-2}%
\end{figure*}

We first consider the low-energy peaks in the resonance density shown in Fig.\ \ref{fig:k-plane-2}.
We solve the effective-medium equation \eqref{eq:avg-green-schrodinger} for the resonance poles in a spherical disordered medium of radius $R$.
The details of this calculation are given in Appendix \ref{app:sphere-scattering} using the partial wave method \cite{Joachain1979, Newton1982, Taylor2006, Sakurai2020}, and the result is the equation
\begin{equation}\label{eq:eff-medium-resonance-2}
\kappa\frac{J_{\nu+1}(\kappa R)}{J_\nu(\kappa R)} = k\frac{H^+_{\nu+1}(kR)}{H^+_\nu(kR)}  \:,
\end{equation}
for a complex resonance at $k\in\mathbb{C}$.
In Eq.\ \eqref{eq:eff-medium-resonance-2}, $\kappa$ is the effective wavenumber \eqref{eq:effective-wavenumber} with $n(\vect{r})=n$, $J_\nu(z)$ and $H^+_\nu(z)$ are the Bessel and Hankel functions, respectively, and the order $\nu$ is given by
\begin{equation}\label{eq:nu-bessel-order}
\nu = \ell + \frac{d-2}{2}  \:,
\end{equation}
where $\ell$ is the orbital quantum number ($\ell\in\{0,1,2,\ldots\}$).
\par The comparison between the numerical solutions of Eq.\ \eqref{eq:eff-medium-resonance-2} for $k\in\mathbb{C}$ and the actual peaks of the resonance density obtained is shown in Fig.\ \ref{fig:k-plane-2}(a) for a 2D point field and in Fig.\ \ref{fig:k-plane-2}(b) for a 3D point field.
In these two panels, instead of Eq.\ \eqref{eq:point-ampli-max}, we use the point scattering model \eqref{eq:point-ampli-hard-sphere} with the scattering length $\alpha=0.1\,\varsigma$ to avoid the divergent behavior of the maximum cross section in the small-$k$ regime.
\par The adequacy between the roots of Eq.\ \eqref{eq:eff-medium-resonance-2} and the peaks in Figs.\ \ref{fig:k-plane-2}(a) and \ref{fig:k-plane-2}(b) is remarkable and supports the validity of the effective-medium equation \eqref{eq:avg-green-schrodinger} to describe the resonance density at large wavelength ($k\varsigma\ll 1$).
It shows that, in this low-energy regime, the incident wave mainly perceives the random point field as a semitransparent sphere without the details due to disorder.
\par
\begin{figure}[ht]%
\includegraphics{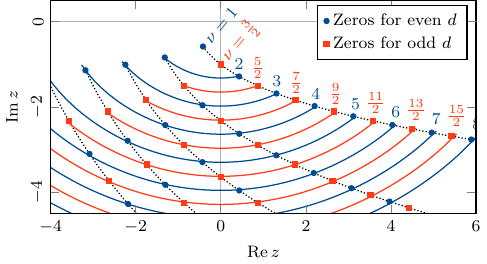}%
\caption{First complex zeros of the Hankel function $H^+_\nu(z)$ for $\arg z\in[-\pi,0]$ at integer and half-integer orders (even and odd dimensions respectively).
The dotted curves represent the trajectory of the zeros when $\nu$ is continuously increased.
Zeros at the same order are connected by a solid line.
The other zeros around $\arg z=-\pi$ for $\nu\in\mathbb{Z}$ are not shown.}%
\label{fig:hankel-zero}%
\end{figure}

 Moreover, it turns out that the resonance structure shown in Fig.\ \ref{fig:k-plane-2} is little affected by the geometric shape of the disordered medium or the scattering parameters such as the cross section of the point scatterers.
Indeed, the same kind of structure can be observed for a cubic medium, instead of a spherical medium.
The robustness of this resonance structure allows us to make a quite radical simplification of the resonance equation \eqref{eq:eff-medium-resonance-2}.
Indeed, given the weak influence of the cross section of the point scatterers, it would be possible to let the parameter $\kappa$ tend to infinity, leading to
\begin{equation}\label{eq:eff-medium-resonance-4}
H^+_\nu(kR) = 0  \:.
\end{equation}
This is the equation of resonances corresponding to the scattering by a \emph{hard sphere} of radius $R$.
The roots of Eq.\ \eqref{eq:eff-medium-resonance-4} are shown in Fig.\ \ref{fig:hankel-zero} and indeed form structures similar to those in Fig.\ \ref{fig:k-plane-2}.
This observation confirms the robustness of this low-energy resonance structure against possibly significant changes of the scattering parameters.
Nevertheless, the hard-sphere resonance equation \eqref{eq:eff-medium-resonance-4} is not as accurate as Eq.\ \eqref{eq:eff-medium-resonance-2} because it completely neglects the scattering within the medium.
This is why the roots of Eq.\ \eqref{eq:eff-medium-resonance-4} are not shown in Fig.\ \ref{fig:k-plane-2}.
In contrast, the prediction of the resonance equation \eqref{eq:eff-medium-resonance-2} of the effective-medium theory is more reliable because it includes the effects of finite values of $\kappa$. %
\par In the 2D case [Fig.\ \ref{fig:k-plane-2}(a)], a slight shift between simulation and prediction can be noticed.
In addition, the predictions in the upper right side of Fig.\ \ref{fig:k-plane-2}(a) do not match the resonance band close to the real $k$ axis.
These disagreements are likely due to multiple-scattering effects because the regime here is diffusive ($R/\lscat\simeq 4$).
Indeed, this would explain why the agreement with Eq.\ \eqref{eq:eff-medium-resonance-2} is better in the 3D case [Fig.\ \ref{fig:k-plane-2}(b)] where the scattering regime is quasiballistic ($R/\lscat\simeq 0.3$).
In order to take into account multiple-scattering effects, we need to go beyond the approach used in this section, which is based solely on the effective-medium equation \eqref{eq:avg-green-schrodinger}.
One possible strategy would be to solve a coupled system formed by Eqs.\ \eqref{eq:avg-green-schrodinger}, \eqref{eq:def-transport-kernel}, and \eqref{eq:transport-eigenvalue-system} for singular values of $k\in\mathbb{C}$.

\subsection{Depth of the resonance band}\label{sec:resonance-band}
Now, we consider the band of resonance poles almost parallel to the real $k$ axis and visible in the upper right corner of Figs.\ \ref{fig:k-plane-2}(a) and \ref{fig:k-plane-2}(b).
In this high-energy region where $\abs{k}\varsigma\gtrsim j_{\frac{d-2}{2}}$, we assume that coherent effects are negligible and that the transport is semiclassical so that we may use the Bethe-Salpeter equation of the form \eqref{eq:transport-eigenvalue-system}.
In order to solve this eigenproblem, we use the diffusion approximation which holds when the mean free path is much smaller than the characteristic size of the disordered medium ($\lscat\ll R$).
As shown in Appendix \ref{app:derivation-diffusion}, the integral equation \eqref{eq:transport-eigenvalue-system} can be cast in this regime into the following diffusion equation and Robin boundary condition:
\begin{equation}\label{eq:diffusion-eigenvalue-system-1}\begin{cases}
\frac{\lscat}{d} \lapl_{\vect{r}}\avg{\rho(\vect{r})} - \gamma \avg{\rho(\vect{r})} = 0  & \text{for}~\vect{r}\in\mathcal{V}  \:,\\[10pt]
\frac{2V_{d-1}}{S_d}\lscat \vect{n}\cdot\grad_{\vect{r}}\avg{\rho(\vect{r})} + \avg{\rho(\vect{r})} = 0  & \text{for}~\vect{r}\in\partial\mathcal{V}  \:.
\end{cases}\end{equation}
The first line of Eq.\ \eqref{eq:diffusion-eigenvalue-system-1} can be written in the more familiar form
\begin{equation}\label{eq:diffusion-variable-separation}
(\lapl_{\vect{r}} + \beta^2)\avg{\rho(\vect{r})} = 0  \:,
\end{equation}
where $\beta$ is the Fourier variable for space, analogous to a wavenumber.
As in Sec.\ \ref{sec:low-energy-resonances}, we assume that the medium is spherical so that the solution of Eq.\ \eqref{eq:diffusion-variable-separation} reads
\begin{equation}\label{eq:diffusion-eigenmode}
\avg{\rho(r)} = j^{(d)}_0(\beta r)  \:,
\end{equation}
where $j^{(d)}_\ell(z)$ is the generalized spherical Bessel function defined in Eq.\ \eqref{eq:def-gen-sph-bessel}.
Using Eq.\ \eqref{eq:diffusion-eigenmode}, the differential problem \eqref{eq:diffusion-eigenvalue-system-1} becomes the following nonlinear system for the pair $(\gamma,\beta)$
\begin{equation}\label{eq:diffusion-eigenvalue-system-2}
\begin{cases}
\frac{\lscat}{d} \beta^2 + \gamma = 0  \:,\\[10pt]
\frac{2V_{d-1}}{S_d}\beta\lscat {j^{(d)}_0}'(\beta R) + j^{(d)}_0(\beta R) = 0  \:.
\end{cases}\end{equation}
The roots of the system \eqref{eq:diffusion-eigenvalue-system-2}, denoted as $\beta_n$ and $\gamma_n$, fully characterize the diffusion eigenmodes in space and time, respectively.
Furthermore, the spectral decomposition of the diffusion problem \eqref{eq:diffusion-eigenvalue-system-1} allows us to write the general solution for the density and its time evolution.
Through inverse Laplace transform, one has the expansion
\begin{equation}\label{eq:diffusion-density-in-time}
\avg{\rho(\vect{r},t)} = \sum_{n=1}^\infty c_n\E^{\gamma_nvt} j^{(d)}_0(\beta_n r)  \:,
\end{equation}
where $v$ is the group velocity and $c_n$ are arbitrary constants depending on the initial condition at $t=0$.
Since the density must be positive, the constant $c_1$, associated with the fundamental mode, should be the largest one.
Note that the values of $\gamma_n$ are negative.
Therefore, Eq.\ \eqref{eq:diffusion-density-in-time} describes the decay of the density in time due to the escape of the particle from the medium.
In the long time limit ($t\rightarrow\infty$), the fundamental mode dominates the other ones, since $\gamma_1$ has the smallest value.
Of course, the superiority of the fundamental mode is required by the constraint on the positivity of the density.
\par Although the boundary condition in Eq.\ \eqref{eq:diffusion-eigenvalue-system-2} cannot be solved for $\beta$ in closed form, it is possible to continue the calculation further if the mean free path is sufficiently small ($\lscat\ll R$).
Indeed, for a small enough $\lscat$, one may recognize in the second line of Eq.\ \eqref{eq:diffusion-eigenvalue-system-2} the first-order expansion of
\begin{equation}\label{eq:diffusion-boundary-sphere-2}
j^{(d)}_0(\beta R_{\rm eff}) = 0  \:,
\end{equation}
where $R_{\rm eff}$ is the effective radius
\begin{equation}\label{eq:diffusion-effective-radius-approx}
R_{\rm eff} = R + \frac{2V_{d-1}}{S_d}\lscat  \:.
\end{equation}
The condition \eqref{eq:diffusion-boundary-sphere-2} obviously expresses the geometrical interpretation of $R_{\rm eff}$ as the radius where the extrapolated density vanishes.
The roots of Eq.\ \eqref{eq:diffusion-boundary-sphere-2} are given by the zeros of $j^{(d)}_0(z)$, or, more explicitly, by the zeros of $J_{\frac{d-2}{2}}(z)$.
Thus, we obtain the full spectrum of diffusion eigenmodes
\begin{equation}\label{eq:diffusion-wavenumber-accurate}
\beta_n = \frac{j_{\frac{d-2}{2},n}}{R_{\rm eff}}  \:,
\end{equation}
where $j_{\nu,n}~\forall n\in\{1,2,3,\ldots\}$ denotes the $n$-th zero of the Bessel function $J_\nu(z)$.
The first zeros of interest are given in Tab.~\ref{tab:bessel-zeros} for the lowest dimensions.
\begin{table}[ht]%
\centering\renewcommand{\arraystretch}{1.3}%
\begin{tabular}{*{5}{c}}\toprule%
$d$   & {\bf 1} & {\bf 2} & {\bf 3} & {\bf 4} \\ \midrule
$n=1$ & $\tfrac{\pi}{2}$  & $2.40483\ldots$ & $\pi$  & $3.83171\ldots$ \\
$n=2$ & $\tfrac{3\pi}{2}$ & $5.52008\ldots$ & $2\pi$ & $7.01559\ldots$ \\
$n=3$ & $\tfrac{5\pi}{2}$ & $8.65373\ldots$ & $3\pi$ & $10.1735\ldots$ \\
$n=4$ & $\tfrac{7\pi}{2}$ & $11.7915\ldots$ & $4\pi$ & $13.3237\ldots$ \\
$n=5$ & $\tfrac{9\pi}{2}$ & $14.9309\ldots$ & $5\pi$ & $16.4706\ldots$ \\ \bottomrule
\end{tabular}%
\caption{First zeros of the Bessel function $J_{\frac{d-2}{2}}(z)$, denoted as $j_{\frac{d-2}{2},n}$ in the text \cite{Olver2010}.
No closed form exists for these zeros, except for $d=1$ and $d=3$.}%
\label{tab:bessel-zeros}%
\end{table}%
According to the first line of Eq.\ \eqref{eq:diffusion-eigenvalue-system-2}, the values of $\gamma$ corresponding to Eq.\ \eqref{eq:diffusion-wavenumber-accurate} read
\begin{equation}\label{eq:diffusion-rate-accurate}
\gamma_n = -\frac{\lscat}{d} \left(\frac{j_{\frac{d-2}{2},n}}{R_{\rm eff}}\right)^2  \:.
\end{equation}
The result \eqref{eq:diffusion-rate-accurate} turns out to be an accurate approximation of the sought eigenvalues of Eq.\ \eqref{eq:transport-eigenvalue-system}.
The quantity \eqref{eq:diffusion-rate-accurate} is also related to the escape rate of the particle per unit time by $\Gamma_n=\abs{\gamma_n}v$ with $n=1$.
\par Finally, using the relationship \eqref{eq:def-gamma-from-k}, the diffusion rate of Eq.\ \eqref{eq:diffusion-rate-accurate} translates into the estimate
\begin{equation}\label{eq:imag-k-accurate}
k_{{\rm idiff},n} = -\frac{\lscat}{2d} \left(\frac{j_{\frac{d-2}{2},n}}{R_{\rm eff}}\right)^2  \:,
\end{equation}
with $n=1$ for the position of the resonance band.
The result \eqref{eq:imag-k-accurate} is considerably more accurate than our previous estimate, Eq.\ (68) of Ref.\ \cite{GaspardD2022b}, since it accounts for the Robin boundary condition \eqref{eq:diffusion-eigenvalue-system-2} which is more precise than the Dirichlet condition $j^{(d)}_\ell(\beta R)=0$. %
Equation \eqref{eq:imag-k-accurate} is also more general than our previous estimate because it predicts other diffusion eigenmodes for $n>1$ than the fundamental one.
\par
\begin{figure}[ht]%
\includegraphics{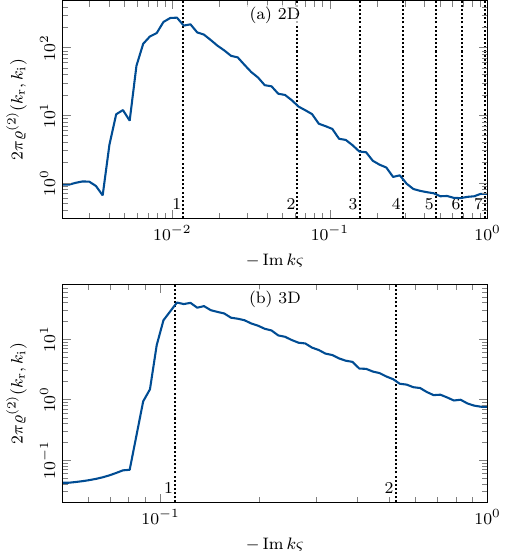}%
\caption{(a) Cross-sectional view of the resonance density \eqref{eq:def-resonance-density} along the vertical axis $\freal{k}=6\,\varsigma^{-1}$ for a 2D disk-shaped random point field with $N=500$ and using the point scattering model \eqref{eq:point-ampli-max}.
The parameters are thus $R=12.6\,\varsigma$ and $\lscat=1.5\,\varsigma$.
The density is averaged over $2^{10}$ disorder configurations.
The vertical dotted lines indicate the diffusion eigenmodes given by Eq.\ \eqref{eq:diffusion-eigenvalue-system-2}.
(b) Same as panel~(a) but in the 3D case with $R=4.92\,\varsigma$ and $\lscat=2.865\,\varsigma$.}%
\label{fig:k-plane-cut-diffusion}%
\end{figure}

 The comparison between the prediction \eqref{eq:imag-k-accurate} and the numerical results is shown in Fig.\ \ref{fig:k-plane-cut-diffusion} for the 2D and 3D cases.
The resonance density (solid curve) is computed with the program MSModel \cite{GaspardD2023-prog} exploiting the resonance potential method of our previous paper \cite{GaspardD2022b}.
Note that the final increase of the density in Fig.\ \ref{fig:k-plane-cut-diffusion}(a) near $\fimag{k}=1\,\varsigma^{-1}$ is due to numerical round-off errors in computing the determinant of $\matr{M}(k)$ and is thus not physical.
This phenomenon was discussed in our previous paper \cite{GaspardD2022b}.
\par The adequacy between the fundamental diffusion mode ($n=1$) and the maximum of the resonance band in Fig.\ \ref{fig:k-plane-cut-diffusion} is remarkable.
This is especially true in 3D, because the scale parameter $n\sigma R=1.72$ is not much larger than $1$ compared to what the diffusion approximation could have required.
In fact, the quality of the approximation \eqref{eq:imag-k-accurate} greatly depends on the expression of $R_{\rm eff}$ and on the boundary condition.
Indeed, if the magnitudes of $R$ and $\lscat$ were the same, then there would be a factor of two between $R$ and $R_{\rm eff}$, and thus a factor of four in $k_{{\rm idiff},n}$, which is significant.
This shows that the accuracy of the prediction \eqref{eq:imag-k-accurate} is not obvious and relies on the value \eqref{eq:diffusion-effective-radius-approx} of the effective radius $R_{\rm eff}$.
\par Noticeably, the other diffusion eigenmodes do not lead to any perceptible structure in Fig.\ \ref{fig:k-plane-cut-diffusion}.
We could have thought that Eq.\ \eqref{eq:imag-k-accurate} predicts the existence of additional bands deeper in the complex plane.
However, we did not find any match, not even in much larger point fields, up to $N=5000$. Only the fundamental mode ($n=1$) appears to play a significant role.
Furthermore, regarding the 1D case, these higher-order modes do no coincide with the faint bands in Fig.\ 6(a) of our previous paper \cite{GaspardD2022b}.
All these observations lead us to conjecture that these higher-order diffusion eigenmodes do not exist in the quantum case.
In this regard, we note that similar distributions, which are devoid of the influence of these extra modes, have been observed in the literature \cite{Kottos2005, Weiss2006, Fyodorov2015, Fyodorov2024}.
\par A likely reason of this absence is that the transport equation \eqref{eq:bethe-salpeter-diag}, on which Eq.\ \eqref{eq:imag-k-accurate} is based, only describes the incoherent (phase-independent) contribution to wave transport.
Therefore, this equation is unable to predict the full resonance distribution in the complex $k$ plane, because the latter would be intrinsically determined by coherence and constructive interferences in the same way as the energy eigenstates in quantum mechanics.
Accordingly, the coherent contributions would smooth out the comb of higher-order diffusion eigenmodes in Fig.\ \ref{fig:k-plane-cut-diffusion}.
\par In principle, the resonance distribution could be determined with advanced methods capable of describing the coherent contributions to wave transport such as the nonlinear sigma model \cite{Efetov1997}.
This field-theoretical method is based on the average of some generating function over the realizations of a random potential.
Upon averaging, the linear wave equation in the disordered medium is then cast into a disorder-free nonlinear wave equation which can be solved semiclassically.
This method has proven particularly effective in determining the full statistics of certain observables \cite{Efetov1997, Nazarov2009, Fyodorov2024}.

\section{Conclusion and perspectives}\label{sec:conclusions}
In this paper, we studied the Foldy-Lax model to describe the propagation of a quantum particle in a disordered medium such as a gaseous detector.
In this model, the atoms composing the medium are represented by point scatterers pinned in space at random positions.
We studied the properties of this system under the perspective of the resonance poles in the complex plane of the wavenumber $k$.
We developed equations for disorder-averaged quantities, in particular the wavefunction and the distribution of complex resonances.
These equations were compared to numerical simulations performed with the program MSModel \cite{GaspardD2023-prog}.
\par More specifically, we showed in Sec.\ \ref{sec:low-energy-resonances} that the peaks in the resonance density at low energy ($\abs{k}\varsigma\lesssim j_{\frac{d-2}{2}}$) are predicted by the effective-medium equation for the disorder-averaged wavefunction.
This means that, when the particle wavelength is larger than the mean interscatterer distance, the wave only probes the geometric shape of the medium and not the details due to the disorder.
\par Finally, in Sec.\ \ref{sec:resonance-band}, we solved the Bethe-Salpeter equation in the diffusion regime to estimate the location of the resonance band at high energy ($\abs{k}\varsigma\gtrsim j_{\frac{d-2}{2}}$).
We showed that the fundamental diffusion eigenmode precisely corresponds to the imaginary part of the resonance band, thus confirming that this band is related to the escape time of the particle in the medium.
However, the higher-order diffusion modes given by the diffusion equation do not lead to visible structures in the resonance density.
A likely reason is that the Bethe-Salpeter equation ignores the coherent effects that should be taken into account to describe the full shape of the resonance density.
In the future, we plan to use advanced methods accounting for coherent contributions to wave transport, such as the nonlinear sigma model \cite{Efetov1997, Nazarov2009, Fyodorov2024}, in order to finely characterize the distribution of resonance widths in strongly open systems.

\begin{acknowledgments}%
The present results were obtained during D.G.'s doctoral thesis defended at ULB in June 2022.
The authors are grateful to Pierre Gaspard for useful discussions and for reviewing this manuscript.
The authors also thank Arthur Goetschy and Romain Pierrat for their support in this research.
D.G. is grateful to Arthur Goetschy for bringing Refs.\ \cite{Efetov1997, Nazarov2009} to his attention.
This work was supported by the Belgian National Fund for Scientific Research (F.R.S.-FNRS) as part of the ``Research Fellow'' (ASP - Aspirant) fellowship program.
This work was also supported by the F.R.S.-FNRS as part of the Institut Interuniversitaire des Sciences Nucléaires (IISN) under Grant No.\ 4.45.10.08.
It also received funding from the French National Research Agency (ANR) as part of the ``MARS LIGHT'' project.
\end{acknowledgments}%

\appendix%
\section{Scattering by a sphere and resonances}\label{app:sphere-scattering}
In this appendix, we derive the formula \eqref{eq:eff-medium-resonance-2} for the complex resonance poles corresponding to the scattering by a spherical effective medium.
The wave equation is Eq.\ \eqref{eq:avg-green-schrodinger} with the spherical-shaped density
\begin{equation}\label{eq:spherical-density}
n(\vect{r}) = \begin{cases}
n  & \text{if}~\norm{\vect{r}} \leq R  \:,\\
0  & \text{otherwise}  \:,
\end{cases}\end{equation}
where $R$ is the radius of the disordered medium.
The effective wavenumber in the region $\norm{\vect{r}}\leq R$ is thus given by Eq.\ \eqref{eq:effective-wavenumber}.
In order to solve Eq.\ \eqref{eq:avg-green-schrodinger} given the spherical symmetry of Eq.\ \eqref{eq:spherical-density}, we will use the partial wave method developed in Sec.\ 1.2 of Ref.\ \cite{GaspardD2022-thesis}.
In this method, the average wavefunction can be expanded in partial waves as
\begin{equation}\label{eq:avg-wfun-pw-expansion}
\avg{\psi(\vect{r})} = \sum_{\ell=0}^\infty \avg{\psi_\ell(r)} P^{(d)}_\ell(\cos\theta)  \:,
\end{equation}
where $P^{(d)}_\ell(\cos\theta)$ denotes the Gegenbauer polynomials generalizing the Legendre polynomials to arbitrary dimension $d$ and defined by
\begin{equation}\label{eq:def-gegenbauer-p}
P^{(d)}_\ell(x) = \frac{\Gamma(\ell + d - 2)}{\ell!\,\Gamma(d-1)} \hypf{-\ell}{\ell+d-2}{\frac{d-1}{2}}{\frac{1-x}{2}}  \:,
\end{equation}
where $\hypf{a}{b}{c}{z}$ is the Gauss hypergeometric function \cite{Olver2010}.
In Eq.\ \eqref{eq:avg-wfun-pw-expansion}, $\cos\theta=\vect{\Omega}\cdot\vect{\Omega}_0$ is the cosine of the angle of $\vect{\Omega}=\vect{r}/r$ with respect to the direction $\vect{\Omega}_0$ of the incident plane wave.
The radial components of the average wavefunction in Eq.\ \eqref{eq:avg-wfun-pw-expansion} satisfy the radial Schrödinger equation
\begin{equation}\label{eq:avg-wfun-radial-equation}\begin{split}
 \bigg[ \der[2]{}{r} + \frac{d-1}{r}\der{}{r} + k^2 & - \frac{\ell(\ell+d-2)}{r^2}  \\
 & - n(r)F(k) \bigg] \avg{\psi_\ell(r)} = 0  \:,
\end{split}\end{equation}
where the density is given by Eq.\ \eqref{eq:spherical-density} for a spherical-shaped medium of radius $R$.
According to this expression of the density, we have to split the wavefunction $\tavg{\psi_\ell(r)}$ in two regions $r\leq R$ and $r>R$.
In this regard, we consider the following ansatz
\begin{equation}\label{eq:eff-medium-radial-wfun}
\avg{\psi_\ell(r)} = \begin{cases}%
j^{(d)}_\ell(\kappa r)   & \text{for}~r\leq R  \:,\\
S_\ell(k) h^{+(d)}_\ell(kr) + h^{-(d)}_\ell(kr)  & \text{for}~r > R  \:,
\end{cases}\end{equation}
where $\kappa=\sqrt{k^2-nF(k)}$ and the generalized spherical Bessel functions are defined by
\begin{equation}\label{eq:def-gen-sph-bessel}\begin{aligned}
j^{(d)}_\ell(z) & = \Gamma(\tfrac{d}{2}) \left(\frac{2}{z}\right)^{\frac{d-2}{2}} J_{\ell+\frac{d-2}{2}}(z)  \:,\\
h^{\pm(d)}_\ell(z) & = \Gamma(\tfrac{d}{2}) \left(\frac{2}{z}\right)^{\frac{d-2}{2}} H^{\pm}_{\ell+\frac{d-2}{2}}(z)  \:,
\end{aligned}\end{equation}
in terms of the standard Bessel function $J_\nu(z)$ and the standard Hankel functions $H^\pm_\nu(z)$ \cite{Olver2010}.
The function $j^{(d)}_\ell(z)$ is finite and regular at $z=0$, hence its presence in the inner region of the ansatz \eqref{eq:eff-medium-radial-wfun}.
Imposing the continuity condition on the wavefunction and its derivative at the boundary $r=R$, we get the following expression for the scattering matrix element
\begin{equation}\label{eq:eff-medium-s-matrix}
S_\ell(k) = -\frac{\Wr[h^{-(d)}_\ell(kr), j^{(d)}_\ell(\kappa r)]_{r=R}}{\Wr[h^{+(d)}_\ell(kr), j^{(d)}_\ell(\kappa r)]_{r=R}}  \:,
\end{equation}
where $\Wr[f(r), g(r)]=f(r)\partial_rg(r)-g(r)\partial_rf(r)$ denotes the Wronskian with respect to $r$.
\par We look for the resonance poles in the scattering matrix element \eqref{eq:eff-medium-s-matrix} coming from the cancellation of the denominator
\begin{equation}\label{eq:eff-medium-resonance-1}
\Wr[h^{+(d)}_\ell(kr), j^{(d)}_\ell(\kappa r)]_{r=R} = 0  \:.
\end{equation}
Finally, expressing $j^{(d)}_\ell(z)$ and $h^{+(d)}_\ell(z)$ in terms of the standard Bessel and Hankel functions, $J_{\nu}(z)$ and $H^+_{\nu}(z)$, we retrieve the resonance equation \eqref{eq:eff-medium-resonance-2}.

\section{Diffusion approximation}\label{app:derivation-diffusion}
One way to obtain the diffusion equation \eqref{eq:diffusion-eigenvalue-system-1} consists in estimating the convolution integral in Eq.\ \eqref{eq:bethe-salpeter-diag}
\begin{equation}\label{eq:transport-convolution-1}
K*(n\sigma\avg{\rho}) = n\sigma \int_{\mathcal{V}} K(\vect{r}\mid\vect{r}') \avg{\rho(\vect{r}')} \D\vect{r}'  \:,
\end{equation}
where $*$ denotes the convolution product and $\mathcal{V}$ the volume of the medium.

\paragraph{Diffusion in the bulk}
We first look at the bulk of the medium, that is to say, far away from the boundary. %
In this region, it is reasonable to extend the integral to $\mathbb{R}^d$, hence neglecting the possible edge effects.
We will consider the special case of the boundary equation later.
Omitting the constant prefactor $n\sigma$, the convolution integral \eqref{eq:transport-convolution-1} reads
\begin{equation}\label{eq:transport-convolution-2}
K*\avg{\rho} = \int_{\mathbb{R}^d} K(\vect{r}\mid\vect{r}') \avg{\rho(\vect{r}')} \D\vect{r}'  \:.
\end{equation}
On length scales much larger than the mean free path ($r\gg\lscat$), an excellent approximation of the kernel $K(\vect{r}\mid\vect{r}')$ is
\begin{equation}\label{eq:transport-kernel-approx}
K(\vect{r}\mid\vect{r}') \simeq \frac{\E^{-(\gamma+n\sigma)\norm{\vect{r}-\vect{r}'}}}{S_d\norm{\vect{r}-\vect{r}'}^{d-1}}  \:,
\end{equation}
according to Eq.\ \eqref{eq:transport-kernel-asym}.
Since $n\sigma$ is very large, the kernel $K(\vect{r}\mid\vect{r}')$ looks like the Dirac delta function $\delta(\vect{r}-\vect{r}')$ plus a correction involving the nonzero variance of the kernel \eqref{eq:transport-kernel-approx}.
To exploit this feature, we may expand the density $\avg{\rho(\vect{r}')}$ in the neighborhood of the observation point $\vect{r}$
\begin{equation}\label{eq:density-series-expansion}\begin{split}
\avg{\rho(\vect{r}')} & = \avg{\rho(\vect{r})} + (r'_i-r_i)\pder{}{r_i}\avg{\rho(\vect{r})}  \\
 & + \frac{1}{2!}(r'_i-r_i)(r'_j-r_j)\pder{}{r_i}\pder{}{r_j}\avg{\rho(\vect{r})} + \cdots \:,
\end{split}\end{equation}
where we have assumed the implicit summation of repeated indices.
We suppose that the first three terms of Eq.\ \eqref{eq:density-series-expansion} will suffice to approach $\avg{\rho(\vect{r}')}$.
The convolution integral \eqref{eq:transport-convolution-2} then amounts to evaluating the first moments of the distribution $K(\vect{r}\mid\vect{r}')$.
The zero-order moment reads
\begin{equation}\label{eq:transport-kernel-moment-0}
\int_{\mathbb{R}^d} \frac{\E^{-(\gamma+n\sigma)\norm{\vect{r}-\vect{r}'}}}{S_d\norm{\vect{r}-\vect{r}'}^{d-1}} \D\vect{r}' = \frac{1}{\gamma+n\sigma}  \:.
\end{equation}
This can be easily shown using the translational and rotational invariances of the integral and the fact that $\D\vect{r}'=S_d(r')^{d-1}\D r'$.
The remaining integral is just $\textstyle\int_0^\infty \E^{-\beta x}\D x=\frac{1}{\beta}$.
The first-order moment
\begin{equation}\label{eq:transport-kernel-moment-1}
\int_{\mathbb{R}^d} \frac{\E^{-(\gamma+n\sigma)\norm{\vect{r}-\vect{r}'}}}{S_d\norm{\vect{r}-\vect{r}'}^{d-1}} (\vect{r}'-\vect{r}) \D\vect{r}' = \vect{0}  \:,
\end{equation}
is equal to the zero vector, because the distribution is symmetric with respect to the point $\vect{r}'=\vect{r}$.
This also means that the particle does not undergo any drift or external forces.
In fact, all the odd-order moments are zero for the same reason.
The second-order moment, which is a $3\times 3$ matrix, reads
\begin{equation}\label{eq:transport-kernel-moment-2}
\int_{\mathbb{R}^d} \frac{\E^{-(\gamma+n\sigma)\norm{\vect{r}-\vect{r}'}}}{S_d\norm{\vect{r}-\vect{r}'}^{d-1}} (r'_i-r_i)(r'_j-r_j) \D\vect{r}' = \frac{2!}{(\gamma+n\sigma)^3} \frac{\delta_{ij}}{d}  \:.
\end{equation}
The off-diagonal elements ($i\neq j$) are zero because of the spherical symmetry of the distribution.
In addition, the three diagonal elements ($i=j$) are equal for the same reason.
These elements can be calculated in spherical coordinates, letting $r_i=r\cos\theta$, but we do not show it here explicitly.
Combining Eqs.\ \eqref{eq:transport-kernel-moment-0}, \eqref{eq:transport-kernel-moment-1} and \eqref{eq:transport-kernel-moment-2}, the convolution integral \eqref{eq:transport-convolution-2} is approached by
\begin{equation}\label{eq:transport-convolution-approx}
K*\avg{\rho} \simeq \frac{\avg{\rho(\vect{r})}}{\gamma+n\sigma} + \frac{\lapl_{\vect{r}}\avg{\rho(\vect{r})}}{d(\gamma+n\sigma)^3}  \:.
\end{equation}
The integral transport equation \eqref{eq:bethe-salpeter-diag} then becomes
\begin{equation}\label{eq:hyp-diffusion-laplace-alt}
\avg{\rho(\vect{r})} = K(\vect{r}\mid\vect{r}_0) + n\sigma\left(\frac{\avg{\rho(\vect{r})}}{\gamma+n\sigma} + \frac{\lapl_{\vect{r}}\avg{\rho(\vect{r})}}{d(\gamma+n\sigma)^3}\right)  \:.
\end{equation}
Multiplying both sides of Eq.\ \eqref{eq:hyp-diffusion-laplace-alt} by $(\gamma+n\sigma)$ readily leads to
\begin{equation}\label{eq:hyp-diffusion-laplace}
\gamma\avg{\rho(\vect{r})} = (\gamma+n\sigma)K(\vect{r}\mid\vect{r}_0) + \frac{n\sigma}{d(\gamma+n\sigma)^2} \lapl_{\vect{r}}\avg{\rho(\vect{r})}   \:.
\end{equation}
Note that Eq.\ \eqref{eq:hyp-diffusion-laplace} is not exactly the standard diffusion equation because the Laplace variable $\gamma$, which can be interpreted as a time derivative ($\gamma\leftrightarrow\partial_{vt}$), is not only in the left-hand side, but also in the right-hand side.
However, since we are only interested in time scales much longer than the intercollisional time ($t\gg (n\sigma v)^{-1}$), the variable $\gamma$ is much smaller than $n\sigma$ and can be neglected in the right-hand side of Eq.\ \eqref{eq:hyp-diffusion-laplace}.
Then, using the property $(\gamma+n\sigma)K(\vect{r}\mid\vect{r}_0)\simeq\delta(\vect{r}-\vect{r}_0)$ in the diffusive regime, Eq.\ \eqref{eq:hyp-diffusion-laplace} becomes
\begin{equation}\label{eq:diffusion-eq-laplace}
\gamma\avg{\rho(\vect{r},\gamma)} = \delta(\vect{r}-\vect{r}_0) + \frac{1}{dn\sigma} \lapl_{\vect{r}}\avg{\rho(\vect{r},\gamma)}  \:,
\end{equation}
whence the first line of Eq.\ \eqref{eq:diffusion-eigenvalue-system-1} since $\lscat=\tfrac{1}{n\sigma}$. %

\paragraph{Leakage boundary conditions}%
Aside from the equation in the bulk, in a finite medium, we also need a condition on the density at the boundary to describe the fact that the wave can freely exit the medium.
Since the medium is much larger than the mean free path, it can be approached by a semi-infinite region
\begin{equation}\label{eq:medium-boundary-plane}
\mathcal{V} \simeq \{\vect{r}\in\mathbb{R}^d \mid \vect{r}\cdot\vect{n}\leq 0\}  \:,
\end{equation}
where $\vect{n}$ is the outward-pointing normal vector to the boundary of the medium.
This assumption is motivated by the fact that, at the scale of the mean free path, the boundary is perceived as nearly flat.
In this way, the boundary is just an infinite plane of equation $\vect{r}\cdot\vect{n}=0$.
In order to obtain a boundary condition on $\avg{\rho(\vect{r})}$, we will approximate the integral transport equation \eqref{eq:bethe-salpeter-diag} for any point $\vect{r}$ on the boundary.
More specifically, we want to estimate the integral
\begin{equation}\label{eq:transport-convolution-edge}
K*\avg{\rho} = \int_{\vect{r}'\cdot\vect{n}\leq 0} K(\vect{r}\mid\vect{r}') \avg{\rho(\vect{r}')} \D\vect{r}'  \:,
\end{equation}
where the integration domain covers the semi-infinite region \eqref{eq:medium-boundary-plane}.
As done in Eq.\ \eqref{eq:density-series-expansion}, we expand the density $\avg{\rho(\vect{r}')}$ in Taylor series at $\vect{r}'=\vect{r}$
\begin{equation}\label{eq:density-expansion-edge}
\avg{\rho(\vect{r}')} = \avg{\rho(\vect{r})} + (\vect{r}'-\vect{r})\cdot\grad_{\vect{r}}\avg{\rho(\vect{r})} + \cdots  \:.
\end{equation}
We neglect the higher-order terms in Eq.\ \eqref{eq:density-expansion-edge}, because we do not want the boundary condition to possess a second-order derivative.
Indeed, the diffusion equation \eqref{eq:hyp-diffusion-laplace-alt} is of second order in space.
Note that, in Eq.\ \eqref{eq:transport-convolution-edge}, the observation point $\vect{r}$ is supposed to lie on the boundary, i.e., to satisfy $\vect{r}\cdot\vect{n}=0$.
Without loss of generality, we can set $\vect{r}=\vect{0}$ due to the translational symmetry of Eq.\ \eqref{eq:transport-convolution-edge} in the plane of the boundary.
The zero-order moment of the kernel reads
\begin{equation}\label{eq:transport-edge-moment-0}\begin{split}
\int_{\vect{r}'\cdot\vect{n}\leq 0} K(\vect{0}\mid\vect{r}') \D\vect{r}' 
 & = \int_{\vect{r}'\cdot\vect{n}\leq 0} \frac{\E^{-(\gamma+n\sigma)r'}}{S_d(r')^{d-1}} \D\vect{r}'  \\
 & = \frac{1}{2(\gamma+n\sigma)}  \:.
\end{split}\end{equation}
This is just half the result \eqref{eq:transport-kernel-moment-0} since it is integrated over half the space.
The first-order moment
\begin{equation}\label{eq:transport-edge-moment-1-eq1}
\int_{\vect{r}'\cdot\vect{n}\leq 0} \vect{r}' K(\vect{0}\mid\vect{r}') \D\vect{r}' = \int_{\vect{r}'\cdot\vect{n}\leq 0} \vect{r}' \frac{\E^{-(\gamma+n\sigma)r'}}{S_d(r')^{d-1}} \D\vect{r}'  \:,
\end{equation}
requires more caution because it is not spherically symmetric. Thus, we need to resort to spherical coordinates.
Using the differential volume element in spherical coordinates
\begin{equation}\label{eq:spherical-diff-volume}
\D\vect{r} = S_{d-1}(\sin\theta)^{d-2} r^{d-1} \D\theta\D r  \:,
\end{equation}
and projecting over the normal vector $\vect{n}$, Eq.\ \eqref{eq:transport-edge-moment-1-eq1} becomes
\begin{equation}\label{eq:transport-edge-moment-1-eq2}\begin{split}
 & \int_{\vect{r}'\cdot\vect{n}\leq 0} (\vect{r}'\cdot\vect{n}) K(\vect{0}\mid\vect{r}') \D\vect{r}'  \\
 & = \frac{S_{d-1}}{S_d} \int_{\frac{\pi}{2}}^{\pi} \D\theta\, \cos\theta(\sin\theta)^{d-2} \int_0^\infty \D r'\, r' \E^{-(\gamma+n\sigma)r'}  \:.
\end{split}\end{equation}
In Eq.\ \eqref{eq:transport-edge-moment-1-eq2}, $\theta$ denotes the angle between $\vect{r}'$ and the normal $\vect{n}$.
The radial and angular integrals of Eq.\ \eqref{eq:transport-edge-moment-1-eq2} are elementary, leading to
\begin{equation}\label{eq:transport-edge-moment-1-eq3}
\int_{\vect{r}'\cdot\vect{n}\leq 0} (\vect{r}'\cdot\vect{n}) K(\vect{0}\mid\vect{r}') \D\vect{r}' = \frac{S_{d-1}}{S_d} \frac{-1}{d-1} \frac{1}{(\gamma+n\sigma)^2}  \:.
\end{equation}
Restoring the arbitrary direction of the normal $\vect{r}$ and using the fact that $V_d=S_d/d$, we finally obtain from Eq.\ \eqref{eq:transport-edge-moment-1-eq3}
\begin{equation}\label{eq:transport-edge-moment-1-result}
\int_{\vect{r}'\cdot\vect{n}\leq 0} \vect{r}' K(\vect{0}\mid\vect{r}') \D\vect{r}' = -\vect{n} \frac{V_{d-1}}{S_d} \frac{1}{(\gamma+n\sigma)^2}  \:.
\end{equation}
Due to the rotational symmetry of the integral in the plane of the boundary, there is no component of the first-order moment \eqref{eq:transport-edge-moment-1-result} parallel to the boundary.
In addition, note the minus sign in the right-hand side of Eq.\ \eqref{eq:transport-edge-moment-1-result} which is a direct consequence of the integration domain $\vect{r}'\cdot\vect{n}\leq 0$.
Combining Eqs.\ \eqref{eq:transport-edge-moment-0} and \eqref{eq:transport-edge-moment-1-result}, the convolution integral \eqref{eq:transport-convolution-edge} can be approximated by
\begin{equation}\label{eq:transport-convolution-edge-approx}\begin{split}
 & \int_{\vect{r}'\cdot\vect{n}\leq 0} K(\vect{r}\mid\vect{r}') \avg{\rho(\vect{r}')} \D\vect{r}'  \\
 & \simeq \frac{\avg{\rho(\vect{r})}}{2(\gamma+n\sigma)} - \frac{V_{d-1}}{S_d} \frac{\vect{n}\cdot\grad_{\vect{r}}\avg{\rho(\vect{r})}}{(\gamma+n\sigma)^2}  \:.
\end{split}\end{equation}
Now, we can substitute the approximation \eqref{eq:transport-convolution-edge-approx} into the integral equation \eqref{eq:bethe-salpeter-diag}.
The source term $K(\vect{r}\mid\vect{r}_0)$ can be neglected, because we assume that it is located far away from the boundary, typically somewhere in the bulk.
This gives us
\begin{equation}\label{eq:transport-equation-edge}
\avg{\rho(\vect{r})} = n\sigma\left(\frac{\avg{\rho(\vect{r})}}{2(\gamma+n\sigma)} - \frac{V_{d-1}}{S_d} \frac{\vect{n}\cdot\grad_{\vect{r}}\avg{\rho(\vect{r})}}{(\gamma+n\sigma)^2}\right)  \:,
\end{equation}
which, after some rearrangements, leads to
\begin{equation}\label{eq:transport-boundary-condition}
\left(\gamma+\frac{n\sigma}{2}\right)\avg{\rho(\vect{r})} + \frac{V_{d-1}}{S_d} \frac{n\sigma}{\gamma+n\sigma} \vect{n}\cdot\grad_{\vect{r}}\avg{\rho(\vect{r})} = 0  \:.
\end{equation}
When considering times much longer than the intercollisional time, the variable $\gamma$ in Eq.\ \eqref{eq:transport-boundary-condition} can be neglected in front of $n\sigma$.
In this way, we obtain a version of Eq.\ \eqref{eq:transport-boundary-condition} which no longer depends on $\gamma$, i.e., giving the time-independent boundary condition
\begin{equation}\label{eq:transport-boundary-condition-approx}
\frac{n\sigma}{2}\avg{\rho(\vect{r})} + \frac{V_{d-1}}{S_d} \vect{n}\cdot\grad_{\vect{r}}\avg{\rho(\vect{r})} = 0  \:,
\end{equation}
on the density $\avg{\rho(\vect{r})}$ for all $\vect{r}\in\partial\mathcal{V}$.
This is the Robin boundary condition given in the second line of Eq.\ \eqref{eq:diffusion-eigenvalue-system-1}.  
Since the coefficients of the two terms in Eq.\ \eqref{eq:transport-boundary-condition-approx} are positive, this condition expresses the fact that the density has an inward-facing gradient.
In the far diffusive regime ($n\sigma R\rightarrow\infty$), this condition reduces to a simple Dirichlet boundary condition: $\avg{\rho(\vect{r})}=0$.
Although this regime is valid for a very large medium, we do not consider this extreme case in this paper.

\end{document}